# Bonds and Bytes: the Odyssey of Structural Biology

S. Hoff[1,†], M. Zinke[2,†], N. Izadi-Pruneyre[2,*], M. Bonomi[1,*]

[1]Institut Pasteur, Université Paris Cité, CNRS UMR 3528, Structural Bioinformatics Unit, Paris, France

[2]Institut Pasteur, Université Paris Cité, CNRS UMR 3528, Bacterial Transmembrane Systems Unit, Paris, France

[†]These authors contributed equally to this work

*Corresponding authors: nadia.izadi@pasteur.fr, mbonomi@pasteur.fr

## Abstract

Characterizing structural and dynamic properties of proteins and large macromolecular assemblies is crucial to understand the molecular mechanisms underlying biological functions. In the field of Structural Biology, no single method comprehensively reveals the behavior of biological systems across various spatio-temporal scales. Instead, we have a versatile toolkit of techniques, each contributing a piece to the overall puzzle. Integrative Structural Biology combines different techniques to create accurate and precise multi-scale models that expand our understanding of complex biological systems. This review outlines recent advancements in computational and experimental methods in Structural Biology, with special focus on recent Artificial Intelligence techniques, emphasizes integrative approaches that combine different types of data for precise spatio-temporal modeling, and provides an outlook into future directions of this field.

**Introduction**

Proteins are key players in the cell as they perform a variety of crucial functions, such as maintaining cell shape and organization, responding to external stimuli, transporting molecules, catalyzing biochemical reactions, regulating gene expression, and defending against pathogens, viruses, and other foreign threats. To perform many of these functions, proteins do not act alone, but in the context of large macromolecular assemblies. Determining the three-dimensional (3D) structure of these complex architectures is a key step to decipher the molecular mechanisms driving biological functions, unravel the origin of dysfunctional behaviors, and design effective strategies to treat associated diseases. However, determining only a static 3D structure is hardly ever enough to fully understand how macromolecular assemblies work. The reason for this is that biological functions often involve dynamic processes that rely on the ability of proteins to interconvert between multiple conformational states [1]. Even disordered regions or proteins that populate a continuum of different conformations can play a crucial role [2], for example in modulating activity and interactions, assisting folding and assembling of supra-molecular complexes, displaying post-translational modification sites, and promoting liquid-liquid phase separation. In order to shed light on biological functions, Structural Biology should therefore aim at determining conformational ensembles that capture structural as well as dynamic properties of biological systems.

There is currently no single experimental or computational technique that can alone generate an accurate, high-resolution, and dynamic description of macromolecular assemblies across multiple spatial and temporal scales. We have instead a Swiss army knife of different approaches, each one providing a single piece of the puzzle: atomistic structures of individual proteins and relatively small biological complexes, descriptions of atomic motions up to the millisecond time scale, medium-low resolution static as well as dynamic images of larger assemblies spanning longer time scales and in the cellular environment (Fig. 1). In this context, Integrative Structural Biology approaches are powerful tools to put all these pieces of the puzzle together into coherent multi-scale models of complex biological systems. In this review, we will first present a brief overview of the recent computational and experimental techniques that have been expanding our knowledge of the structural and dynamic properties of proteins and macromolecular assemblies. We will then focus on integrative approaches that can leverage all this information in a synergistic way to build accurate spatio-temporal models of biological systems. Finally, we will discuss possible future directions in Structural Biology and try to delineate what we should expect in this field in the next 10 years.

**Main text of review**

*Computational approaches in Structural Biology*

Structural Biology has a long tradition of computational techniques to predict protein structure from its amino acid sequence, ranging from classical comparative approaches based on homologs of known structure [3] to ab-initio fragment-based modeling [4,5], often guided by spatial restraints derived from the analysis of sequence co-evolution [6,7]. Furthermore, over the years Molecular Dynamics (MD) simulations have provided precious insights into the dynamics of individual proteins and larger complexes at atomistic detail, and more recently up to the entire cell at a more coarse-grained resolution [8]. Continuous developments in the force field used to drive MD simulations as well as enhanced-sampling techniques to accelerate the exploration of the conformational space [9] have enabled the determination of accurate protein structural ensembles across the wide spectrum between order and disorder.

Lately, the field of Structural Biology has experienced a profound transformation due to the AI revolution. While AlphaFold2 [10] (AF2) marked a significant milestone, this revolution is truly the result of years of prior experimental, computational, and technological advancements, as well as the accumulation of precise structural data in the PDB database. Tools such as AF2, RoseTTAFold [11], OmegaFold [12], and OpenFold [13], which enable quick and accurate protein structure prediction from sequence, have already had a monumental impact on the life sciences communities, for example by helping solving X-ray crystallography phase problems [14], generating initial structures to be used for cryo-EM refinement [15], molecular dynamics simulations [16] or virtual screening [17], and more generally for stimulating new hypotheses.

Although the blistering pace at which these novel approaches have been developing makes it difficult to foresee the potential of AI, significant and clear limitations do exist today [18]. First, AF2 and other AI tools are currently unable to predict structures in biologically relevant conditions, such as in crowded cellular environments, nor the effect on protein structure of salt, pH, post-translational modifications, small molecules, and mutations. This issue is further compounded by the limited accuracy of AF2 in predicting the structure of protein complexes. Thus, the determination of protein-protein interactions remains one of the most significant and potentially rewarding challenges. Finally, AI based approaches often give only 'snapshot' views of proteins in their most populated states. However, recent works have focused on extracting thermodynamically relevant conformations from AF2 for proteins that populate a few distinct states [19-21] as well as more disordered systems [22,23]. These efforts suggest that tools such as AF2 contain information on the full protein conformational landscapes which, if properly extracted, may help link protein structure and dynamics to function.

*Experimental approaches in Structural Biology*

Alongside these new developments in the computational area, Structural Biology experimental techniques have also faced tremendous progress. Historically, X-ray crystallography was considered the gold standard for determining high-resolution structures of macromolecules, particularly proteins. However, an X-ray structure represents only a static 'snapshot' of a protein captured in a crystalline state. The recent developments in single-particle cryo-electron microscopy (cryo-EM) have revolutionized the field of Structural Biology by enabling the visualization of macromolecules in near-native conditions and in different conformational states [24]. However, information on dynamics and the transitions between different states is still out of cryo-EM reach.

The power of dynamic structural biology techniques (Fig. 1) in contrast to snapshot techniques lies in their ability to directly study internal dynamics and their role in folding, conformational changes and molecular associations at different time scales. Of these techniques, only NMR spectroscopy can achieve atomic resolution and cover a wide range of time scales (from ps-ns to μs-ms and hours). A prime example is the recent study by the Kern lab [25], where the authors performed exchange-sensitive NMR spectroscopy experiments to map the free-energy landscape of the adenylate kinase. This study provided atomic details of the conformations involved in the catalytic mechanism, especially in the motions of the protein lid domain in the μs timescale, which are accessible at this resolution only by NMR.

The combination of time-resolved approaches with snapshot techniques allows to integrate structural, kinetic, and thermodynamic information, resulting in "movies" that are able to characterize complex and dynamic cellular processes. One recent example by the Kay lab [26] shows how the complementarity of the cryo-EM snapshots of a key component of *Mycobacterium tuberculosis* protein degradation machinery and NMR data allows to unravel its allostery catalytic mechanism at atomic detail. The insight provided by this study will facilitate the design of new inhibitors.

A comprehensive understanding of these processes necessitates the cellular context, which can be achieved by in-cell NMR [27], cryo-electron tomography (cryo-ET) [28], cellular mass spectrometry and light microscopy [29]. For example, the combination of cellular cross-linking mass spectrometry and cryo-ET was used to determine the organisation of an RNA polymerase-ribosome supercomplex in the context of *M. pneumoniae* cell. The integration of both approaches allowed for a subnanometer resolution model where mass spectrometry data was essential to identify a novel protein from an unoccupied electron density [30].

*Integrative modelling approaches*

The synergistic use of different computational and experimental Structural Biology techniques to address specific biological questions has characterized this field for a long time. Inspired by this integrative philosophy, several computational techniques have been developed to combine the information provided by different types of experiments and convert them into accurate and precise structural models. These so-called *integrative modelling* approaches are based on a common architecture [31] (Fig. 2): *i)* the molecular components of a model and their representations are defined by the input information; *ii)* scoring functions are used to quantify the agreement of a model with all the input information; *iii)* a sample of models is generated. Finally, models are analyzed, and the process is usually repeated until the models are deemed sufficiently consistent with the input information and sufficiently precise for addressing the biological questions of interest. Bayesian scoring functions [32] account for expected and unexpected errors in the input information and can be used to integrate different types of experimental data based on their accuracy. More recently, integrative approaches have been developed to model conformationally heterogeneous systems with ensemble-averaged data [33-37], such as SAXS and NMR, often based on Bayesian inference and the Maximum Entropy principle [38,39]. The implementation of these techniques in open-source software, such as IMP [40], ROSETTA [4], PLUMED-ISDB [41], powER [42], Assembline [43], HADDOCK [44], BioEn [34], and BME [35], has accelerated their adoption. Furthermore, structural models obtained with integrative approaches can now be deposited in the PDB-Dev database [45].

A classic example of integrative structure and dynamic determination of complex biological systems is the Nuclear Pore Complex (NPC), a large macromolecular assembly that selectively transports cargoes across the nuclear envelope. Initial low-resolution models of the yeast NPC were obtained in 2007 by integrating a large variety of biophysical and proteomic data [46] and later refined to sub-nanometer precision [47]. More recently, by adding fluorescence correlation spectroscopy calibrated live imaging, an integrative and dynamic model of the postmitotic NPC assembly pathways enabled the identification of intermediate states populated during the assembly process [48]. In parallel to these efforts, the architecture of the human NPC as well as its large-scale dilations *in cellulo* were determined by the Beck lab using an integrative approach that incorporated cryo-ET and light microscopy data with AI-built structural models of the individual NPC components [49].

Another example of integration of complementary approaches to study the structure and dynamics of complex systems is the bacterial pili or endopili of the type II secretion system. These pili are filaments inserted into the cytoplasmic membrane of many bacteria species and composed of protein subunits. Their assembly and disassembly drive the secretion of key virulence factors across the bacterial envelope. In 2004, the first structural model was determined by combining the X-ray structure of the soluble domain of an individual subunit, a model of its transmembrane domain and negative stained EM images of the whole fiber at 2.5 nm resolution [50]. Subsequently, a flexible docking approach that exploited this data along with new mutation and phenotypic assays provided an atomic model of the pilus [51]. In 2017, an NMR study revealed the presence of calcium in the monomer structure, exhibiting a native structural arrangement distinct from the previously determined X-ray structure [52]. The integration of the high-resolution NMR structure of the monomer subunit with the medium-resolution cryo-EM map of the pilus led to a pseudo-atomic model of the pilus at 5 Å resolution with unexpected atomic details. The dynamics of these fibers, key for their functions, was also explored by a combination of experimental and computational approaches (NMR, HDX-MS, and normal mode analysis) on isolated and assembled proteins [53]. Recently, the complementarity between X-ray and NMR data was essential to determine the highly dynamic association mode of proteins involved in the pilus assembly. This information on structure, dynamics and assembly of different proteins of the system, along with their relative cellular abundance and in vivo cross-linking data, were combined to propose a mechanistic model of how the pilus assembly drives secretion [54].

**Conclusions**

This brief overview of the odyssey of Structural Biology was meant to highlight how significant advancements in the field, recently propelled by AI, are making the process of obtaining structures of complex macromolecular architectures and characterizing their dynamic properties easier, faster, cheaper, and more accurate. This leads us to wonder whether the discipline of Structural Biology will soon be dead. The answer is probably yes. At least if we hold onto the idea that the main purpose of this discipline is to obtain a structure or create a movie. With the risk of stating the obvious, this has never been the ultimate goal. On the contrary, structures have always been just the beginning of the exploration of the molecular mechanisms underlying biological functions. As this step is becoming increasingly less challenging, researchers might find themselves with more time to redirect their focus toward the true scope of Structural Biology: understanding how biological systems work.

In the future, we will put more effort into tackling even more complicated questions, exploring larger systems, and investigating multi-step processes occurring across various time scales. We will start looking more closely and at higher resolution at how proteins work in their native cellular environment, where interactions with other partners, like nucleic acids, sugars, lipids, and other cell components, play a central role. We will also focus on dissecting the effects of the cellular environment, such as ions, salt concentration, pH, and crowding, on structural and dynamic properties as well as understanding how post-translational modifications and missense mutations affect biological functions [55]. This is an area where AI techniques have a lot to offer but they have yet to demonstrate their potential. However, with the growing complexity of the questions addressed and the systems studied, there is no doubt that the integration of multiple different *in silico* and experimental approaches will always be needed.

Design will certainly play a crucial role in the near future, if not today. Recent works have already demonstrated the potential of AI for *de novo* design of proteins, antibodies, nanobodies, and small molecules that modulate protein functions [56-58]. These efforts, supported by experimental validation, will be fundamental for the development of new drugs, therapies, and treatments for various diseases. Looking forward, a more difficult task will be understanding how to design molecular tools to modulate the functions of highly dynamic systems, such as disordered proteins and RNA molecules. The challenge here is twofold: first, identifying in complex conformational landscapes which substates are responsible for specific functions or subfunctions and then designing agents that can shift the equilibrium in the desired direction.

When we look at the future of Structural Biology from a broader perspective, it becomes evident that our focus will extend well beyond the study of individual proteins, protein complexes, or even large macromolecular architectures. Instead, we will deal with an interconnected environment in which cascades of interactions and pathways link a multitude of partners to ensure the functioning of cells and high-order organizations, such as tissues and organisms. Pioneering modelling approaches, such as Bayesian metamodeling [59], are starting to bridge the gap between Structural Biology, System Biology, and Cell Biology and delineate a future in which these disciplines converge in a truly *Integrative Biology*. In this scenario, Structural Biology is very much alive and with a bright future ahead.

## Acknowledgements

M.Z. was funded by the French Agence Nationale de la Recherche (ANR Energir ANR-21-CE11-0039). S.E.H. was founded by a Roux-Cantarini fellowship from the Institut Pasteur (Paris, France). The authors thank Andrej Sali for providing the material used to prepare Figure 2 (and its caption) and Falk Schneider for helpful discussions on microcopy techniques.

# List of Figures

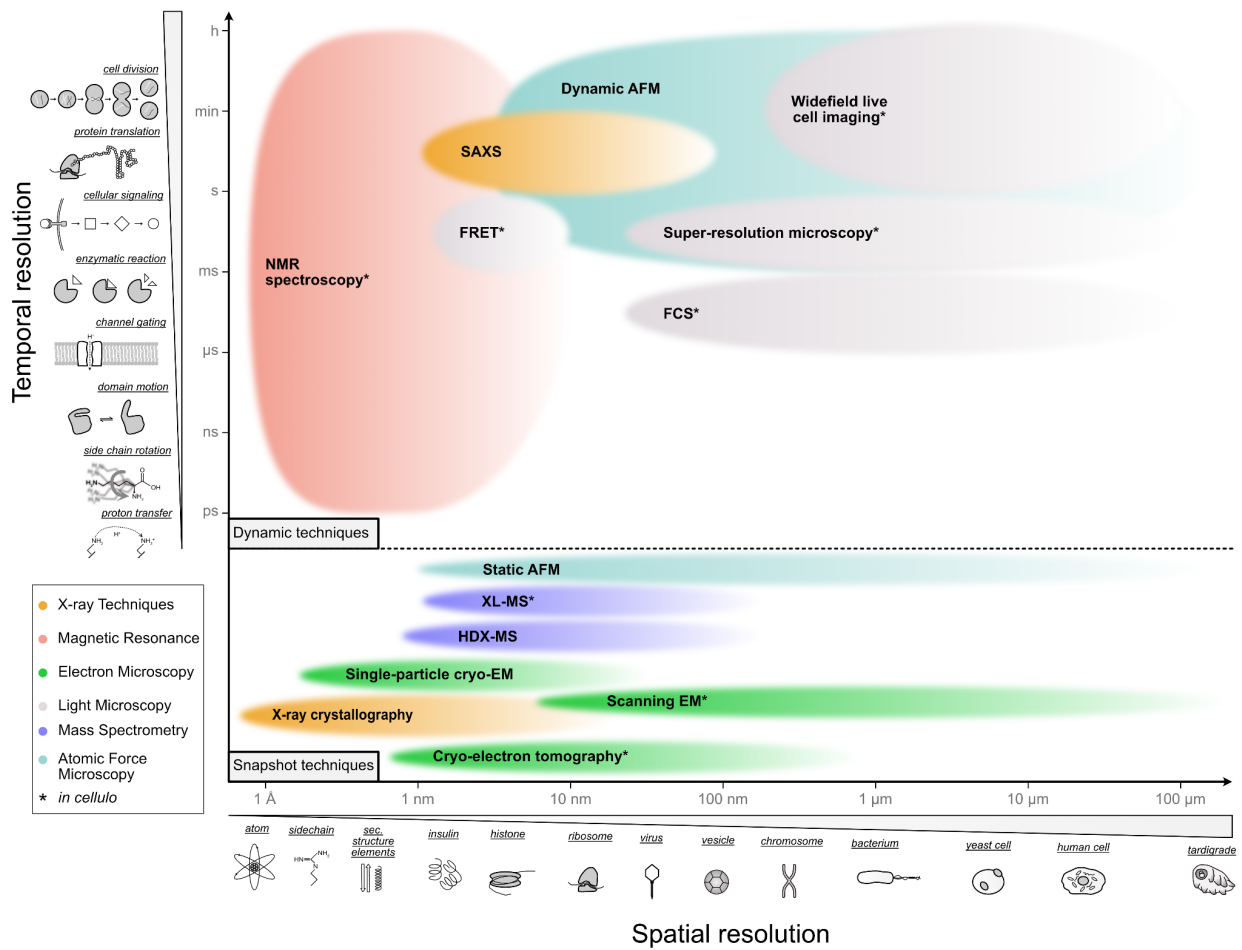

**Figure 1.** *Experimental structural biology: dynamic and snapshot techniques.* Plot of temporal resolution versus spatial resolution, with representative examples for each axis. These examples depict typical magnitudes but are not exclusive to the indicated values. Techniques based on X-rays are represented in orange, those using magnetic resonance in red, electron microscopy methods in green, light microscopy in gray, mass spectrometry in purple, and AFM in blue. Techniques feasible *in cellulo* are marked with an asterisk. Dynamic methods allow for the direct observation of time-resolved motions, whereas in snapshot methods these motions are restrained due to crosslinking, freezing or crystal formation. It has to be noted that in many cases snapshot techniques might allow for the extraction of structural data representing different states, e.g., by capturing different class averages in vitreous ice with cryo-EM. However, kinetic and thermodynamic information regarding this exchange processes remain inaccessible.

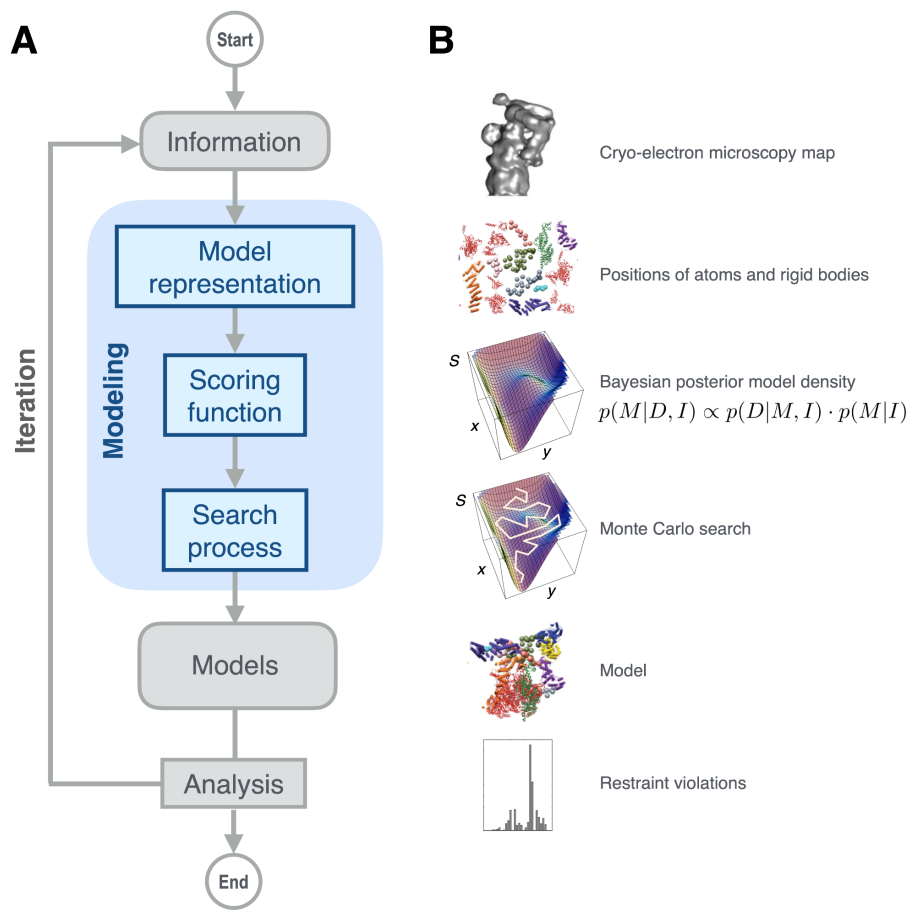

**Figure 2.** *Integrative structure modeling workflow.* **(A)** Integrative structure modeling is an iterative process that converts the input information about a biomolecular system into structural models. The input information specifies the composition of the system in terms of its components and the spatial relationships between them. The scoring function quantifies the agreement of a model with the input information. Once a sufficient sample of acceptable models (model ensemble) is in hand, it is analyzed and the process is potentially repeated, until the model ensemble is deemed sufficiently consistent with input information and sufficiently precise for addressing biological questions of interest. The blue rectangle indicates a single instance of modeling. **(B)** Each aspect of integrative modeling is illustrated with an example. Information is represented by a cryo-electron microscopy density map. The representation is illustrated by a mixture of Cartesian coordinates of individual atoms and rigid bodies. The scoring function S is illustrated by a two-dimensional funnel-like energy landscape. The sampling is represented by a stochastic Monte Carlo search (white line) on the scoring function landscape. A model is illustrated by a single instance of a structure that satisfies the input information. Analysis is represented by a histogram of the number of models as a function of the number of restraint violations per model.